# Magnetic properties of isolated Re ion and Re-Re complex in ZnO studied by GGA +U approach

O. Volnianska [*]

Institute of Physics PAS, al. Lotników 32/46, 02-668 Warsaw, Poland

**Abstract**

Magnetic properties of Re substituting for the divalent Zn ions and impurity pairs in wurtzite and zinc blende ZnO are analyzed by employing the Generalized Gradient Approximation with the +U corrections. The +U term applied to $d$(Zn), $p$(O), and $d$(Re) orbitals stabilizes spin polarization of Re. Re impurity introduces a defect level located in the band gap, and $Re^{2+}$ generates the local spin 5/2 in both wurtzite and zinc blende structures. Magnetic coupling of Re-Re pairs as a function of distance between the defects, their relative orientation, and the charge state was calculated. All magnetic coupling between impurities is antiferromagnetic and due to an intermediate O atom via strong $d$(Re)-$p$(O) hybridization between Re and O states. The defect states and magnetic moments come from $d$(Re) with a contribution of $p$(O) orbitals of the O nearest neighbors of Re. The results show that the Re-doped wurtzite ZnO may be a new type of magneto-optical materials with a great promise.

[*]volnian@ifpan.edu.pl

## 1. Introduction

ZnO becoming dilute magnetic semiconductor (DMS) via partial replacement of the cations by magnetic transition-metal (TM) ions is highly attractive for magneto-optical, magneto-electrical, magneto-transport and spintronic devices [1–5].

Ferromagnetism (FM) in *p*-type Mn-doped ZnO (predicted by T. Dietl *et al.* in 2000 [2]) is one of the most widely studied, but also controversial one due to a number of contradictory results observed for this system. Mn substituting for the divalent Zn in ZnO introduces the level located in the band gap [6, 7], and $Mn^{2+}$ shows the local magnetic moment amounting to 5 $\mu_B$. A wide array of different results were obtained for collective magnetism in ZnO:Mn ranging from FM [8, 9], to antiferromagnetism (AFM), to paramagnetism (PM), and to spin glass phase [10].

The mechanism of FM in TM-doped oxide semiconductors remains unclear, because the observed magnetism can be either due to an intrinsic property of the material, or a result of the formation of secondary magnetic phases coming from precipitation of TM clusters [10]. One of the explanations of magnetism in TM-doped ZnO is based on the first principles calculations showing that the overlap of 3*d*(TM) with *p* orbitals of neighboring O atoms forms delocalized band levels, and an exchange interaction depends on the density of spin polarized states at the Fermi level [11-14]. It was shown that wave function of Mn ion in ZnO is strongly localized on Mn states, and magnetic interaction between Mn-Mn is short ranged [10, 11, 13]. It was pointed out that a short-range exchange coupling is a dominant property in a double exchange or super-exchange interactions [15, 16]. Moreover, most of the reported experimental magnetization values are much smaller than the expected value for a free Mn ion. Delocalization of impurity-induced bands may be responsible for long-range magnetic interactions and stabilization of FM phase via *p-d* exchange interaction [17, 18]. As it was noted, local distortion may play an important role in the magnetic behaviour of DMSs [19].

To check if the above properties hold also for rhenium Re ions in wurtzite (*w*) and zinc blende (*zb*) ZnO, the Density Functional Theory calculations within the generalized gradient approximation (GGA) including the Hubbard-like term +U describing the on-site Coulomb interactions [20] where preformed. Our choice of Re as impurity in ZnO is justified by the facts that (i) 5*d*(Re) orbital, like 3*d*(Mn), is occupied by 5 electrons and can assume substantial magnetic moment, and (ii) atomic radius of Re is larger than one Zn which can lead to substantial atomic relaxation and strong delocalization of the wave functions of the defect states. The delocalization of electrons on the defect states can lead to magnetic interaction between local moments. On the other hand, it needs to be noted, the strong delocalization weakens the stability of local high spin (HS) state of impurity.

Previous theoretical studies of the properties of TM-doped in DMSs revealed that the results strongly depend upon the used exchange-correlation functional [7, 21, 22]. Differences between the LDA/GGA and the GGA+U approaches explain the discrepancies between the energies of band gap ($E_{gap}$) and defect levels, and localization of wave functions. As a rule, the inclusion of the +U correction (or the usage of hybrid functional) increases the localization of states. Moreover, the structures of spin polarized wave functions of Zn vacancy in ZnO as predicted by GGA and GGA+U are qualitatively different [23]. In particular, in Refs. [24, 25] it was obtained using GGA+U that magnetic interactions between rhenium impurities in silicon are FM.

In the present paper we study the isolated Re and Re-Re pairs in *w*-and *zb*-ZnO, and analyze the impact of the +U terms on their properties. After presenting the details of calculations in Sec. 2.1, we apply the +U terms to *d*(Zn) and *p*(O) orbitals, and determine their values by fitting the GGA+U band gap and electronic structure of ZnO to experiment (Sec. 2.2). Next, we present results of calculations of the defects formation energies (Sec. 2.3). We discussed the effect of spin-orbit coupling (SOC) on electronic structure Re in *w*-ZnO in Sec. 2.4. We compare the Heyd-Scuseria-Ernzerhof (HSE) and GGA+U magnetic properties of defects in Sec. 2.5. Sec. 3.1 contains the results of calculations of electronic structure of isolated Re and the impact of +$U_{Re}$ on the magnetic structure of a defect. The magnetic coupling between Re-Re in *w*-and *zb*-ZnO is analyzed in Sec. 3.2. Next, the structural relaxation is discussed in Sec. 4. Finally, Sec. 5 summarizes the results.

## 2. Methods

### 2.1 Methods of calculations

Calculations based on the density-functional theory were performed using the generalized gradient approximation (GGA), [26] the Perdew-Burke-Ernzerhof (PBE) exchange-correlation potential, [27] including the +U term implemented in the QUANTUM-ESPRESSO code [28] along with the theoretical framework as developed in Ref. [20]. Ultrasoft atomic pseudopotentials were employed, and the following valence orbitals were chosen: $3d^{10}$ and $4s^2$ for Zn, $2s^2$ and $2p^4$ for O, $6s^2$, $6p^0$, $5d^5$, $5f^0$ (and $6s^2$, $6p^0$, $5s^2$, $5p^6$, $5d^5$ for full relativistic pseudopotential) for Re. The two above pseudopotentials for Re give rise to the same results. The plane wave basis with the kinetic energy cutoff ($E_{cutt}$) of 40 Ry provided a convergent description of the analyzed properties. The Brillouin zone summations were performed using the Monkhorst-Pack scheme with a 2×2×2 $k$-point mesh [29]. Methfessel-Paxton smearing method with the smearing width of 0.136 eV was used for obtaining partial occupancies. Ionic positions were optimized until the forces acting on ions were smaller than 0.02 eV/Å. The wurtzite 72 and 128-atoms and zinc blende 64 and 216-atoms supercells with one or two Zn atoms replaced by Re ions were used in calculations.

To verify the convergence of the results with the respect to the size of the supercell, test calculations were performed for isolated Re and two defects displaced by ~ 6.5 Å (4NNa configuration in notation of Sec. 3.2) using 192- and 512- atom supercells for $w$- and $zb$-ZnO, respectively. The results showed that energies of spin polarization ($\Delta E^{PM-FM}$) (defined in Sec. 3.1) agree to within 0.15 eV or less, and energies of magnetization ($\Delta E^{AFM-FM}$) (defined in Sec. 3.2) agree to within 0.02 or less. The good convergence is connected with the strong localization of defect states.

## 2.2 Changes in electronic structure of ZnO induced by +U correction

At first, the electronic structure and density of states (DOS) of pristine material were calculated. The +U term was applied to $3d$(Zn) and $2p$(O) orbitals allowing it to vary from 0 to 10 eV. The ionic positions were fixed at ideal sites, and then relaxed. The values of the U terms were fitted to reproduce the band structure of ZnO, in particularly, $E_{gap}$ and the energy of $d$(Zn) state. In fact, the LDA/GGA approach shows smaller $E_{gap}$ [30] and the deep valence $d$(Zn) level higher when comparing to the experimental data [31]. Our analysis finds that $U_{Zn}$ =10 eV and $U_O$ = 7 eV reproduce both the experimental $E_{gap}$, 3.4 eV [32-34] and the energy of the $d$(Zn) band, centered about 8.1 eV below the VBM [28] for $w$- ZnO. Calculated $U_{Zn}$ and $U_O$ are similar to the values reported in other theoretical works [7, 35-37].

Our approach of correcting the $d$(Zn) and $p$(O) states is justified by the fact that deficiencies of the LDA/GGA are of general character, and the underestimation of the gap follows from the sublinear dependence of the LDA/GGA total energy on the occupation [20, 21, 38]. Moreover, the sensitivity of $E_{gap}$ to both $U_{Zn}$ and $U_O$ is explained by the orbital compositions of both the VBM and the minimum of the conduction band (CBM). The calculations showed that the main contributions to the DOS in the region of the VBM are made by $3d$(Zn) and $2p$(O) orbitals states. The calculations with $U_{Zn} = 10$ eV and $U_O = 7$ eV give the lattice constants $a_w$=3.223, $c_w$= 5.24 Å ($a_{zb}$= 4.5 Å), which are very close to the experimental values [34, 39].

## 2.3 Formation energy of defects

To study the property of Re-doped ZnO, we introduce one Re to the system, which corresponds to the concentrations of 1.5, 1.2, and ~1 % for 128, 192 and 216 atom supercells. Formation energy $E_{form}$ of a defect is [40, 41]:

$$E_{form} = E_{tot}(ZnO:\text{Re}) - E_{tot}(ZnO) + \sum n_i \mu_i + q(E_F + E_{VBM}) + E_{correct}, \qquad (1)$$

where the first two terms on the right-hand side are the total energies of the supercell with and without the defect Re$_{Zn}$, respectively, and $n_i=\pm 1$ with the + (−) sign corresponding to the removal (addition) of an atom. $\mu_i$ are the variable chemical potentials of atoms in the solid, which in general are different from the chemical potentials $\mu_i(bulk)$ of the ground state of elements, *i.e.*, Zn *bulk*, Re *bulk* and O$_2$. Chemical potentials of the components in the standard phase are given by the total energies per atom of the elemental solids: $\mu$(Zn *bulk*) = $E_{tot}$(Zn *bulk*), $\mu$(Re *bulk*) = $E_{tot}$(Re *bulk*), while $\mu$(O *bulk*) = $E_{tot}$(O$_2$)/2 that is the total energy per atom for O$_2$. In calculation of the formation energy of Re$_{Zn}$ in O-rich condition, $\mu$(Zn) = $E_{tot}$(Zn *bulk*) + $\Delta H_f$(ZnO) and $\mu$(Re) = $E_{tot}$(Re *bulk*) + $\Delta H_f$(ReO$_3$) are taken, where $\Delta H_f$ is the enthalpy of formation per formula unit, and it is negative for stable compounds. $\Delta H_f$ at T = 0 K is obtained by considering the reaction to form or decompose a crystalline bulk ZnO and ReO$_3$ from or into its components and dependent on an cohesive energy, $E_{coh}$, of Zn, Re, and O. The obtained results for $E_{coh}$ of Zn, Re, O in their equilibrium structures are of 1.32 (1.35 [42]), 8.1 (8.01 [42]), and 2.86 (2.65 [42]) eV, in reasonable agreement with experimental data (shown above in brackets). The calculated and the measured enthalpies formations for the binary oxides ZnO and ReO$_3$ structure are also in reasonable accord, -3.7 (-3.6 [43]) and -3.01 (-2.9 [44]) eV.

The last term, in Eq.1 $E_{correct}$, includes two corrections. The first one, $\Delta E_{PA}$, is the potential alignment correction of the VBM. The VBM in the ideal supercell and in the supercell with a (charged) defect differ by the electrostatic potential, and is obtained by comparing the potential at two reference points far from the defect in the respective supercells with ($P[D^q]$) and without ($P[0]$) the defect, $\Delta E_{PA} = q(P[D^q] - P[0])$. Second correction is an image charge correction as expressed by Makov-Payne form: $E_{MP} = \dfrac{q^2 \alpha_M}{2\varepsilon_d W^{1/3}}$, where $\alpha_M$ is the lattice-dependent Madelung constant, which are 1.64 and 1.75 for *zb* and *w* structures, respectively. *W* is the supercell volume, and $\varepsilon$ is the static dielectric constant of ZnO, equal to 8.5.

The calculated formation energies of the $Re_{Zn}$ are in the Table I. $E_{form}$ in *w*-ZnO is 2.1 eV which is a few times higher than $E_{form}$ of Mn in ZnO [45]. Nevertheless, this is a reasonable value, which corresponds to a concentration of defect of $10^{11}$ cm$^{-3}$.

**2.4 The effect of spin-orbit coupling**

Because Re is the heavy transition metal, the electronic structure of ZnO:Re can therefore be affected by relativistic effects, including spin-orbit coupling. In order to verify the SOC effects on magnetic properties of Re in ZnO the test calculations were performed using fully relativistic pseudopotentials based on the PBE functional [46]. The calculated total energies of previously relaxed *w*-ZnO:Re with GGA+U method indicate that when SOC is taken into account, the total spin moments decrease by 8 % and 3 % for 72- and 128-atom supercells, respectively, due to some decrease of Re spin moment. Moreover, the Re defect states with SOC in comparison with the corresponding ones without SOC show only slightly more split „$t_{2g}$" level (see Sec 3.1 where the detail discussion of electronic structure of ZnO:Re are presented). Thus the spin-orbit coupling was neglected in the current work.

**2.5. GGA+U vs HSE results**

In order to verify the correctness of GGA+U magnetic properties results, the test calculations were performed for the hybrid functional HSE, based on the PBE functional where parameter α is fraction of the exchange that is replaced by Hatree-Fook exchange [47]. Calculations were performed for isolated Re and Re-Re (1NNa configuration in notation of Sec. 3.2) using 72- and 128- atom supercells for *w*- ZnO. Screening parameter α =0.42 was set [48] to correctly reproduce band gap, 3.2 eV. Results show that HSE defect levels are higher by about 0.3 eV relative to valence band maximum (VBM) than GGA +U defect levels due to the more downshift of the VBM for large α. We note that the problem of choosing the α parameter to get accurate defect levels is still open [49,50], as well choosing the U parameter in GGA + U approach. Nevertheless, the results indicate that energies of spin polarization ($\Delta E^{PM-FM}$) (defined in Sec. 3.1) agree to within 0.3 eV or less, and energies of magnetization *($\Delta E^{AFM-FM}$)* (defined in Sec. 3.2) agree to within 0.015 or less. That shows the good agreement for magnetic properties for these two approaches.

## 3. Results

The Section summarizes the obtained results for defect levels, density of states, spin density, and discusses the influence of $U_{Re}$.

### 3.1 Electronic structure of isolated Re

Electronic structure of Re ion in ZnO is determined by its local atomic configuration which assumes a tetrahedral structure. According to a simple molecular orbital model, the positions of $Zn^{2+}$ ions are replaced by $Re^{2+}$ ions, and $Re^{2+}$ ($d^5$ configuration) introduces defect states that can be regarded as combinations of the *d*(TM) with four dangling bonds of "Zn vacancy", *i.e.* with four $sp^3$ orbitals of four O neighbors (see Figures 1a-c). In *zb*-phase, the *d*(TM) shell splits into a doublet $e_g$ and a higher in energy triplet $t_{2g}$. The wurtzite field further splits the $t_{2g}$ into a doublet and a singlet, and this pair is here denoted by "$t_{2g}$". $e_g$ is a localized atomic like-state and "$t_{2g}$" couples with the $t_{2g}$ *p*(O) states creating bonding and antibonding energy levels [51].

### 3.1.1 Influence of the $U_{Re}$ term

Both spin-polarized and nonmagnetic calculations were performed to calculate the total energy, the electronic structure, and the total spin ($S_{tot}$) of ZnO:Re. The stability of spin state was examined by the calculation of the energy of spin polarization $\Delta E^{PM-FM}$ defined as the difference in the total energy of the spin-nonpolarized and spin-polarized phases. $\Delta E^{PM-FM}$ is positive when spin polarization is stable, and the defect in this case is in the high spin state. If $\Delta E^{PM-FM} \leq 0$, the defect is in the low spin (LS) ground state.

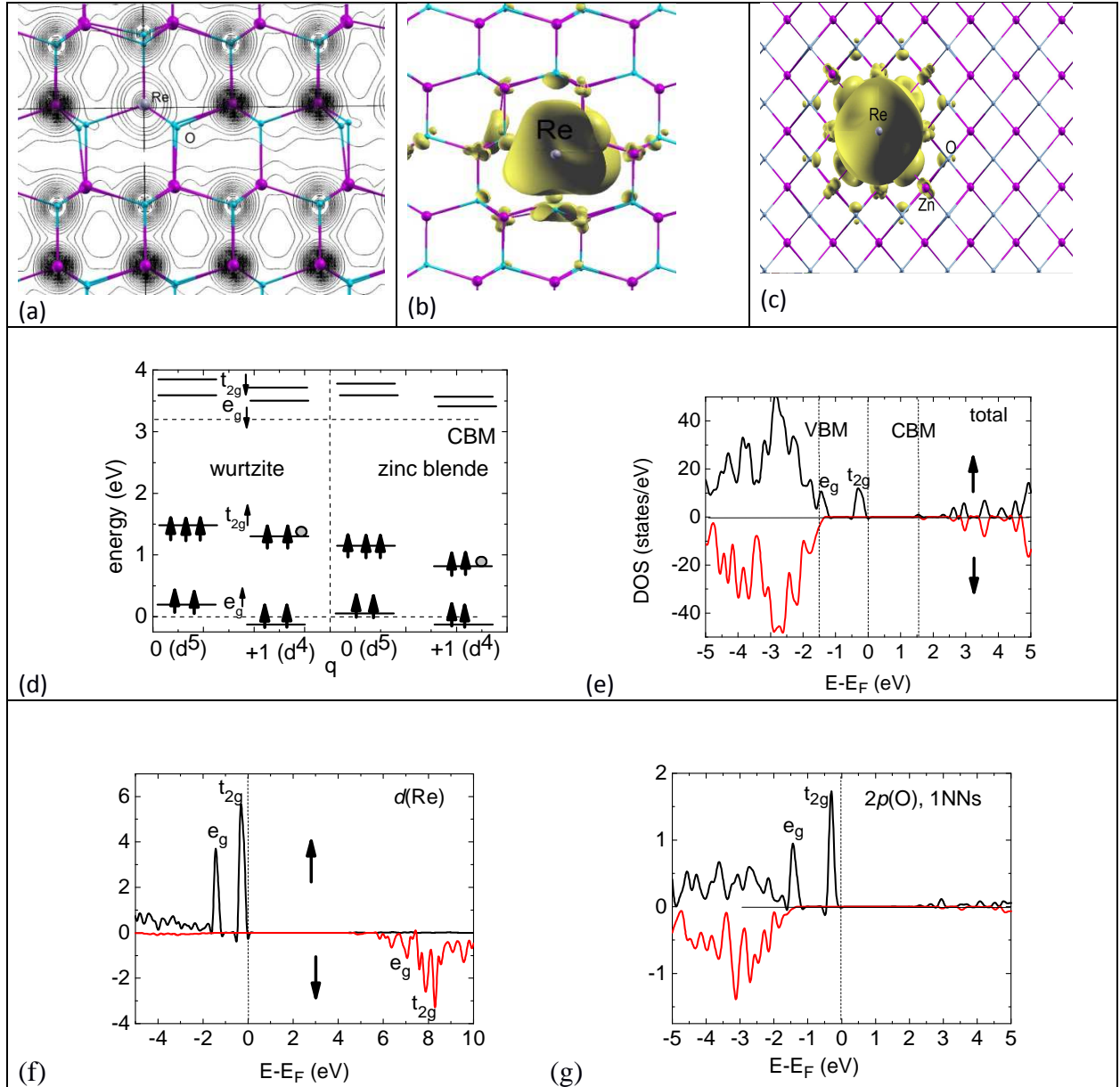

Fig. 1 (a)-(c) Calculated atomic configurations and isosurfaces of: (a) the electron and (b), (c) spin densities corresponding to 0.001 electron/bohr$^3$ in (a), (b) $w$- and (c) $zb$- ZnO:Re. (d) the $e_g\uparrow$, "$t_{2g}\uparrow$" of Re$^{2+}$ and Re$^{3+}$ relative to the VBM; (e)-(g) spin-resolved DOS of $w$-ZnO:Re: (e) total DOS, (f) the contribution of 5$d$(Re) orbitals, (g) the contribution of 2$p$(O) orbitals of the nearest rhenium O neighbors. Negative values denote the spin down channel. $U_{Re}$ = 5 eV.

The calculations show that abovementioned properties are heavily dependent on $U_{Re}$ values ($U_{Re}$ was treated as a free parameter changing from 0 to 8 eV). When $U_{Re}$ < 2 eV, Re is in the LS state with both $S_{tot}$ and the $\Delta E^{PM-FM}$ equal to zero. With the increasing of $U_{Re}$ from 2 to 5 eV, the spin polarization becomes finite, $S_{tot}$ increases from 0 to 5/2, and $\Delta E^{PM\text{-}FM}$ rises to 3.7 eV (see Table I). When $U_{Re} > 5$ eV, $S_{tot} = 5/2$ in all cases, and $\Delta E^{PM\text{-}FM}$ changes only about 0.4 eV. Hence, $U_{Re} = 5$ eV is chosen in this work. The impact of $U_{Re}$ term on the spin polarized properties of Re in ZnO can be attributed with the fact that $U_{Re}$ term affects the localization of the defect states, the localization of wave functions makes the exchange coupling stronger and increases spin polarization energy [21-23, 52].

Re in ZnO introduces a doublet $e_g$ and a higher in energy triplet "$t_{2g}$" states located in the band gap that split by exchange splitting $\Delta\varepsilon_{ex}$ into spin-up ("$t_{2g}\uparrow$", $e_g\uparrow$) and spin-down ("$t_{2g}\downarrow$", $e_g\downarrow$) levels, where $\Delta\varepsilon_{ex(t)} = \varepsilon("t_{2g}\downarrow") - \varepsilon("t_{2g}\uparrow") > 3$ eV and $\Delta\varepsilon_{ex(e)} = \varepsilon(e_g\downarrow) - \varepsilon(e_g\uparrow) > 3$ eV (Fig. 1d). Here, $\varepsilon$ is the energy of the defect level relative to the VBM. "$t_{2g}\uparrow$" and $e_g\uparrow$ (occupied by 5 electrons) located in the band gap about 0.2 and 1.5 eV above the VBM, respectively (Fig. 1d). The defect is in the HS ground state with $\Delta E^{PM\text{-}FM} = 3.7$ eV (see Table I), and shows the magnetic moment amounting to 5 $\mu_B$ located as 2.7 $\mu_B$ on Re ion and 2.3 $\mu_B$ on four nearest neighbor O ions.

Figs. 1b and 1c indicate that Re induced states are dominated by the localized and spin polarized contribution of the $d$(Re) orbital and four $sp^3$ orbitals of the O nearest neighbors. By calculating the contributions of individual atoms projected onto relevant atomic orbitals to the total DOS (Fig. 1e) one finds that the main contribution to the impurity states ($e_g$, "$t_{2g}$") comes equivalently from $5d$(Re) and $2p$(O) of the O nearest neighbors of Re (Figs. 1f and 1h), in agreement with Figs. 1b.

From Table I and Fig. 1d it follows that the results for $w$-ZnO:Re are qualitatively the same with the results for $zb$-ZnO:Re.

TABLE I. The formation energy $E_{form}$, the spin polarization energy $\Delta E^{PM-FM}$, the energy of magnetic coupling $\Delta E^{AFM-FM}$ in eV, together with the total and absolute ($S_{abs}$) spins in the charge state $q$ obtained in $w$- and $zb$- ZnO:Re.

| | | $w$-ZnO | | | $zb$-ZnO | | |
|---|---|---|---|---|---|---|---|
| | | | Re | | | | |
| q | | $E_{form}$ | $\Delta E^{PM-FM}$ | $S_{tot}$ | $E_{form}$ | $\Delta E^{PM-FM}$ | $S_{tot}$ |
| 0 | | 2.1 | 3.7 | 5/2 | 1.95 | 3.0 | 5/2 |
| +1 | | 1.05 | 2.9 | 2 | 1.0 | 1.56 | 2 |
| | | | Re-Re | | | | |
| | | | $\Delta E^{AFM-FM}$ | $S_{tot}/S_{abs}$ | | $\Delta E^{AFM-FM}$ | $S_{tot}/S_{abs}$ |
| 0 | 1NNa | | -0.28 (AFM) | -0.13/4.93 | 1NNa | -0.18(AFM) | 0/4.8 |
| | 1NNc | | -0.21 (AFM) | -0.170/5.0 | - | - | - |
| | 4NNa | | -0.05 (AFM) | -0.1/4.8 | 4NNa | -0.006 (AFM) | 0/4.6 |
| | 4NNc | | -0.02 (AFM) | -0.05/4.9 | - | - | - |
| +2 | 1NNa | | -0.3 (AFM) | -0.2/4.2 | 1NNa | -0.08(AFM) | -0.01/3.9 |
| | 1NNc | | -0.28 (AFM) | -0.19/4.1 | - | - | - |
| | 4NNa | | -0.06 (AFM) | -0.08/4.0 | 4NNa | -0.002 (AFM) | -0.02/3.1 |
| | 4NNc | | -0.052(AFM) | -0.04/4.0 | - | - | - |

### 3.1.2 $Re^{2+}$ vs $Re^{3+}$

$Re^{2+}$ and $Re^{3+}$ that assume the charge state $q=0$ and +1, respectively, are shown in Fig. 1d. The difference in the energy of the levels corresponding to q=0 and +1 are determined by two counteracting effects. The first one is the intracenter Coulomb repulsion, which decreases with the number of electrons and leads to the decrease of the defect levels. The second effect is the dependence of the +U-correction on the occupancy levels and its hybridization with the host states [20]. From Fig. 1d it follows that in both $w$- and $zb$-ZnO the intracenter Coulomb repulsion is larger than the $+U_{Re}$ occupancy effect, because the Re-states in q = +1 are lower in energy with the decreasing occupancy. The increase of q from 0 to +1 leads to the decrease of "$t_{2g}$" of about 0.3 eV with respect to the VBM. In both $w$- and $zb$-ZnO, the defect is in the stable HS state with $\Delta E^{PM-FM}$ = 2.9 and 1.6 eV, respectively, and $S_{tot}$ = 2 (see Table I). The change of the charge state of Re is determined by the transition level between q=0 and q=+1, $\varepsilon(0/+)$. It is defined as the Fermi energy relative to the VBM at which formation energies of the 0 and +1 charge states are equal, q=0 → q=+1:

$$\varepsilon(0/+) = E_{form}(q=0) - E_{form}(q=+1), \qquad (2)$$

where $E_{form}$ is defined by Eq. 1. The calculated values for $w$- and $zb$-ZnO relative to the VBM are 1.05 and 0.95 eV, respectively.

### 3.1.3 Charge and spin density

Figs. 1a-c show isosurfaces of electron (Fig. 1a) and spin (Figs. 1b,c) densities for supercells containing Re. The contour (Fig. 1a) is plotted in the (100) plane and shows that a large contribution to the electron density comes from the oxygen atoms and strong ionic bonds resulting from $p$(O)-$d$(Re) hybridization. The spherical symmetry around oxygen ions takes place, which indicates that the bonds in Re-doped ZnO are dominated by the ionic component.

For U(Re) = 5 eV, the spatial dependence of the spin density (Figs. 1b,c) is determined by two factors: the energies $e_g$ and "$t_{2g}$" of Re relative to the VBM and the symmetry of the crystal. We note that strong localization of the spin density is on Re ion because "$t_{2g}$" is a deep gap state. The spin density of Re decays exponentially but comprises also a long-range tail which involves $p$(O) orbitals of distant O ions due to the strong $p$(O)-$d$(Re) hybridization. Both $p$(O) and $d$(Re) orbitals build also the VBM of ZnO (see Figs. 1e-g). The contribution of the $d$(Zn) orbitals to the spin density is non-negligible due to the substantial contribution of $d$(Zn) to the VBM. Secondly, spin polarization in wurtzite structure exhibits an anisotropic character as seen in Fig. 1b. In $zb$-ZnO, spin density is isotropic (Fig. 1c).

### 3.2 Magnetic interaction of Re-Re pairs

It was shown above that +U correction applied to $d$(Zn), $p$(O) and $d$(Re) orbitals stabilizes spin polarization of Re in ZnO, but it does not guarantee that magnetic coupling between Re-Re is also stabilized. This is because a stronger localization of the defect wave function may reduce the coupling, as it was obtained for vacancies in ZnO [23, 41].

The electronic structure of Re-Re pair is determined by: (i) the distance between ions, (ii) the relative orientations of Re ions with respect to each other and to the crystal axes, and (iii) the charge state. In $w$-ZnO four configurations were considered, Re-Re as the first (1NN) and the fourth (4NN) nearest neighbors in the same (*x, y*) basal plane perpendicular to the *c*-axis (referred here as the *a* case), and oriented along the *c*-axis (denoted here as the *c* case). The spatial separations in the 1NNa, 1NNc, 4NNa, 4NNc cases are close to 3.22 (2.93), 3.21 (3.0), 6.45 (6.41) and 6.15 (6.13) Å, respectively (the relaxed distances are shown in the brackets). In *zb*-ZnO, 1NNa and 4NNa are 3.27 (3.11) and 6.56 (6.53) Å, respectively. Finally, we mention that whereas the Re-Re pair in the 4NNs configurations has eight nearest neighbor O atoms, both the 1NNa and the 1NNc have only seven nearest neighbors, because one O atom is shared by two Re.

### 3.2.1 Magnetization energy of Re-Re

When Re-Re is spin polarized, the orientation of the two spins can be either FM or AFM (the total spin ($S_{tot}$) vanishes, but $\Delta E^{PM-FM}$ is finite). The magnetization energy $\Delta E^{AFM-FM}$ is defined as a difference between the total energies of AFM and FM states, and is negative when AFM coupling is stable. The calculated $\Delta E^{AFM-FM}$ are summarized in Table I. In both *w*- and *zb*- structures 1NN pairs are spin polarized the magnetic ground state is AFM, and it does not depend on the charge state. Moreover, the absolute value of the coupling strength is high, about $0.2 - 0.3$ eV and absolute spin about 5 and 4, for q=0 and +2, respectively. The dependence of the coupling on the orientation in *w*-ZnO is rather weak: the total energy differences between the *a*- and *c*-orientations are smaller than $\Delta E^{AFM-FM}$. In contrast to *zb*-ZnO where Re-Re demonstrates short-ranged magnetic interaction, in wurtzite crystal $\Delta E^{AFM-FM}$ for 4NNa is relatively high, about $0.05-0.06$ eV (see Table I). In *zb*-ZnO, $\Delta E^{AFM-FM}$ for 4NNa configuration is only 0.002-0.005 eV which indicates that AFM coupling is slightly more stable than FM coupling, and magnetic interaction is very weak.

### 3.2.2 Mechanism of magnetic coupling Re-Re

As it was pointed out above, the simple molecular-like model showing that a Re ion induces a doublet-triplet pair is corroborated in the detailed GGA+U calculations. In the following scheme we extend this simple model to the case of two interacting Re ions. Here, two pairs ($e_g$, "$t_{2g}$") Re$_1$ and ($e_g$, "$t_{2g}$") Re$_2$ are coupled (via the same spin states) and form bonding (B) and antibonding (A) combinations (see Fig. 2a). Figs. 2a,b show FM and AFM mechanisms of magnetic interaction of Re-Re. The dominant mechanism is determined by the interplay between the magnitude of the B-A splitting and the exchange of spin-up – spin-down splitting. Because the highest occupied $t_{2g}$ state is fully occupied, the energy gained from the super-exchange interaction is larger than the one gained from the double-exchange interaction. This is due to the substantial B-A splitting. As a consequence the AFM phase is privileged. With the increase of distance between the two Re ions, the AFM coupling of $d$(Re)-$d$(Re) state becomes weaker, and the doped system can exhibit PM phase, which is consistent with the results from Table I.

To further understand the origin of magnetism of Re in the ZnO, Figs. 2 c-f show the density of $5d$(Re) and $2p$(O) states of the 1NNa Re-Re pair in the AFM phase. There is an asymmetry between spin-up and spin-down DOS and non-vanishing spin polarization on the Fermi level. The exchange splitting between $d$(Re)-$d$(Re) and $d$(Re)-$p$(O) may be observed from Figs. 2c-e and it can be seen that bonding $e_g^B$ and $t_{2g}^B$ states are fully occupied and antibonding states are empty. Bandwidth of B-A splitting is large due to strong hybridization effects. Because the Re ion has $d^5$ electronic configuration, the spin-up $d$ orbitals of Re are all occupied, which favors the AFM configuration (see Fig. 2b). In contrast to Mn and other $3d$(TMs), which assume FM coupling in $p$ type host, the charged defect Re-Re pair in q=+2 demonstrates AFM behavior due to the very large B-A splitting.

Figs. 2g-i show spin densities calculated for several Re-Re configurations in $w$- and $zb$-ZnO. The figure reflects the shape and localization of the occupied $e_g$ and $t_{2g}$ states of the Re-Re complexes. These states are localized on the $d$(Re) orbitals and $p$ orbitals of the seven O atoms (see Figs. 2g,i) or on the eight neighbors for more distant pairs (see Fig. 2h). In the wurtzite crystal the spin densities are more delocalized than in $zb$ structure, since they comprise long-range tails involving $p$ orbitals of distant O ions by broken crystal symmetry. In $zb$-ZnO all O neighbors of the complex are almost equivalent, as displayed by the similar contributions to the spin density. A comparison of those pictures shows that the orbitals of agiven Re ion are spin-polarized in the opposite way. In case of the 1NN pair, the two Re ions share one O ion, the $p$ orbitals of which contribute to the states of both defects. This "orbital frustration" leads to a repopulation between the $sp^3$ spin-orbitals of all the O ions (Figs. 2f,g,i)

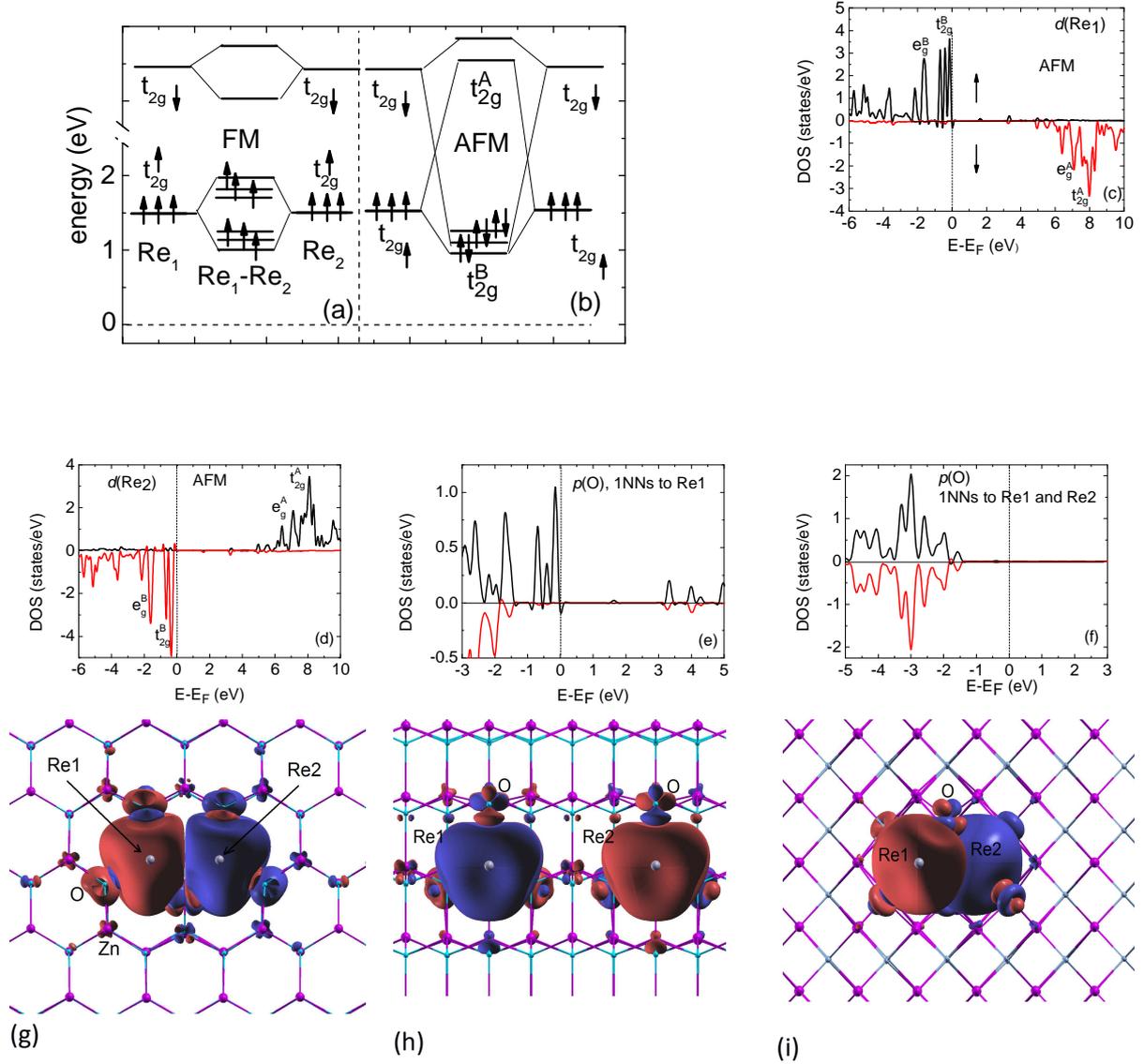

Fig. 2 (a), (b) The scheme of ferromagnetic and antiferromagnetic coupling mechanisms in *w*-ZnO. (a) FM, (b) AFM. (c)-(f) DOS of ZnO:Re-Re in AFM state; (c) the contribution of 5*d*(Re) orbitals of the first Re ion, (d) the contribution of 5*d*(Re) orbitals of second Re ion, (e) the contribution of 2*p*(O) orbitals of the nearest to rhenium O neighbors, (f) the contribution of the nearest to rhenium O' neighbors, O' is shared by both Re ions. Negative values denote the spin down channel. (g)-(i) Calculated isosurfaces of spin densities corresponding to 0.001 electron/bohr$^3$. (g) 1NNa is shown in *xy* plane, *w*-ZnO. (h) 4NNa is shown in *xz* plane, *w*-ZnO. (i) 1NNa is shown in *xy* plane, *zb*-ZnO. Blue and red color correspond to different spin orientations. Spin density of a given defect are presented from different directions for more clarity.

## 4. Distortion in Re-doped ZnO

As it can be seen in Figs. 1a-c and Figs. 2g-i, significant perturbations in the crystal structure of ZnO are introduced by the large radius of Re. In particular, we find $d_a$(Re-O) =2.26 Å and $d_c$(Re-O) = 2.29 Å, which are larger by about 15-16 % than corresponding Zn-O bonds in *w*-ZnO. (The bonds in the same (*x, y*) basal plane are referred as *a* subscript, and bonds oriented along the *c*-axis denoted with *c* index). Moreover, in *zb*-ZnO the Re-O bond is larger than Zn-O by about 13 %. Because of pseudo $C_v$ symmetry of Re in *w*-ZnO only three basal O atoms nearest to Re are equivalent. From the results of Table II it follows that distortion in the *c*-direction is smaller than in the basal plane. Generally, it can be noted that the structural relaxation in ZnO:Re is a few times larger than in case of ZnO:Mn [7].

TABLE II. Zn-O, Re-O, and Re-Re bond lengths in ZnO:Re and ZnO:Re-Re. Re-O' denotes the bond length between Re and O' atom which is shared by two Re. All values are in Å.

|           | *w*-ZnO   |           | *zb*-ZnO  |
|-----------|-----------|-----------|-----------|
|           | $d_a$     | $d_c$     | $d_a$     |
| Re        |           |           |           |
| Zn-O      | 1.95      | 2.01      | 2.04      |
| Re-O      | 2.26      | 2.29      | 2.25      |
| Re-Re     |           |           |           |
| Re-Re     | 2.92      | 3.04      | 3.11      |
| Re-O/Re-O'| 2.26/2.19 | 2.29/2.21 | 2.25/2.22 |

Also for Re-Re pair in ZnO we find large structural relaxation. Actually, in *w*-ZnO $d_a$(Re-Re) =2.92 Å, which is smaller than the ideal Zn-Zn bond length (3.22 Å). In *zb*-ZnO $d_a$(Re-Re) is 3.11 Å in contrast to $d_a$(Zn-Zn) which is 3.27 Å. Re-O and Zn-O bond lengths are also different by about 15 %. Here, we note that a detailed description of the structural distortions is not straightforward, because the atomic relaxations around Re and Re-Re involve not only the nearest but also more distant neighbors. As it was noted in Ref. [19] the difference in the magnetic properties of Re impurity in *w*- and *zb*-ZnO can be connected with lattice distortion, which can play a crucial role in magnetic behavior.

### 5. Summary and conclusions

Spin states of Re and magnetic coupling between Re-Re in *w*- and *zb*-ZnO were studied within GGA+U calculations. The $U_{Zn}$ = 10 eV and $U_O$ = 7 eV terms were imposed on the *d*(Zn) and *p*(O) orbitals, leading to the correct band gap of ZnO. For U(Re) < 2 eV, spin of Re is equal to zero. The inclusion of the $U_{Re}$ = 5 eV term strongly enhances spin polarization of Re states. As a result, the S=5/2 and S=2 high spin configurations are stabilized for the respective q=0 and q=+1 charge states. Magnetic moments originate mainly from Re 5*d* orbitals and the contribution of *p* orbitals of oxygen is about 40 %. The results indicate that AFM coupling is the most stable for four doping configurations with two charged states, namely 0 and +2. Re-Re magnetic coupling of the nearest neighbor pairs is -0.28 eV, while the coupling of 4NN pairs is -0.005 eV. Thanks to that Re-doped *w*-ZnO may be promising magnetic semiconductor.

## ACKNOWLEDGEMENTS

The work was supported by FNP grant POMOST/2012-5/10 and the National Science Centre, (Poland) Grant No. 2015/17/D/ST3/00971. Calculations were done at Interdisciplinary Center for Mathematical and Computational Modeling, University of Warsaw (Grant No. GA65-27). I would like to thank Prof. P. Boguslawski for reading the manuscript of this paper and making remarks.